\documentstyle[12pt]{article}
\normalbaselines
\begin{document}

\begin{center}
\vspace*{1.0cm}
{\LARGE {\bf Evaporating Black Holes And An Entropic Scale-Hierarchy\\ }}

\vskip 1.5cm

{\large {\bf Hadi Salehi}}

\vskip 0.5 cm

Institute for Studies in Theoretical Physics and
Mathematics, \\ P.O.Box 19395-1795, Niavaran-Tehran, Iran
\\ and\\
Arnold Sommerfeld Institute for Mathematical Physics, TU Clausthal, \\
Leibnizstr. 10, D-38678 Clausthal-Zellerfeld, Federal
Republic of Germany
  \end{center}

\vspace{1cm}

\begin{abstract}
It is argued that a characteristic length may be associated with
the entropic state of a spherically symmetric
black hole in the cosmological context. This length is much smaller than
the Schwarzschild-radius of a black hole and
may act as a regulator of arbitrarily
high frequencies apparently entering
the usual derivation of Hawking's radiation.
\end{abstract}
\vspace{1cm}
\section{Introduction}
One of the most impressive predictions of general relativity
is that the process of gravitational contraction of a 'sufficiently' large
mass at
some stages leads to the formation of a black hole. For an outside
observer a black hole appears to be a state in which the maximal
achievable limit of entropy
production by gravitational contraction is achieved. The form of entropy
function was suggested by Bekenstein [1] by means of information
theoretic arguments. His result was greatly strengthened by Hawking's
remarkable discovery [2] that quantum effects lead to the association of a
radiation with the black hole mass at a temperature which corresponds
to the assumed form of information theoretic Bekenstein entropy, namely
($c$ resp. $\hbar$ is the velocity of light resp. the
Planck constant)
\begin{equation}
T_H= \frac{\hbar c^3}{4\pi M G},
\label{I1} \end{equation}
where $M$ denotes the black hole mass.\\
Perhaps it is one of the most important characteristics of this formula
that it interconnects all the fundamental constants of the nature in a
single thermodynamic relation and it is widely believed that
a generic understanding of this interconnection
will lead to new fundamental
insights about the unification of physical concepts in general. \\
The present paper, in essence, is the outgrowth of attempts to understand
the origin of this
interconnection from a quite universal point of view. It should, however,
be remarked
at the outset that many
aspects of the ideas to be reported
are far from being dynamically
justified and the problem of a strict  dynamical approach is not
addressed.
Accordingly, the paper ought to be understood as
an extremely modest contribution towards a strict dynamical
elaboration of the results.

\section{Cosmic thermalization of gravitational constant}

The idea as to whether the gravitational constant might
fundamentally be related to
an entropic property of the universe as a whole to my
knowledge has never been worked out in any details. A strong indication
suggesting an entropic approach is provided by the relation
(\ref{I1}). When written in the alternative form
\begin{equation}
G^{-1}=4\pi \frac{M T_H}{\hbar c^3},
\label{c1}\end{equation}
it suggests that in the presence of a black hole an outside asymptotic
observer may interpret the gravitational constant
as an emergent feature of an equilibrium type process
having its origin in the
gravitational trend of a 'sufficiently' large mass towards a maximal
entropic state. There is, of course, an inherent ambiguity in a statement
of this type, for it is impossible to give an account of the entropic
state of a sufficiently large mass without an extrinsic knowledge of the
numerical value of $G$. \\
One may consider two different feasible ways
of establishing the value of gravitational constant extrinsically.
In fact, one might either simply identify
the 'extrinsic' value of $G$, denoted
in the following by $G_{ex.}$ as a constant of nature, or else consider it as
fundamentally related to $\hbar$, $c$, the temperature of the
microwave background radiation $T_{mic.}$ and the entire visible mass
of the universe $M_{uni.}$\footnote{For this mass I shall take
in the following the value $10^{53}kg$ with the realization that
there may exist possible uncertainties in the estimate of this value}.
Although these are two alternative ways
of establishing the value of $G_{ex.}$, in the presence of a black
hole the physical implications in the second case will be very different.
In the rest of this paper an attempt is made to explore the physical
implications of the second alternative.\\
First, it should be realized that in a universe which is thermodynamically
unique there
must be only one feasible way
of relating the extrinsic value of
$G$ to $\hbar$, $c$, $T_{mic.}$ and $M_{uni.}$.
Quite in the sense of (\ref{c1}) this relation may be assumed to be
linear with respect to the equilibrium parameter (temperature), namely
\begin{equation}
G_{ex}^{-1}=f(\hbar,c,M_{uni.},\alpha)~T_{mic.} ,
\label{c2}\end{equation}
where $\alpha$ is imagined to represent a still unspecified
quantity needed to establish the value of $G_{ex.}$. The explicit form
of the relation (\ref{c2}) may be suggested
by a general consideration. First,
note that (\ref{c2}) would act as a stringent constraint imposed
by the state of the universe as a whole on the locally observed value of
physical constants about an arbitrary point of observation. Since such a
constraint would affect rules of local measurements, so by
necessity $\alpha$ must be treated as a 'cosmic' field $\alpha(x)$.\\
It should, however, be realized that, not knowing absolut standards of units
in the process of local measurements,
the conceivability of a cosmic field
$\alpha(x)$
regulating the thermodynamic connection between
all physical constants
about an arbitrary point of observation is meaningless unless $\alpha(x)$
is invariant under a position
dependent transformation of units. The need for this limitation
is apparent as the value of $\alpha(x)$ at each space time point is to
reflect an absolute property of the universe as a whole and, hence, must
remain unaffected by a change of the particular standards
of units used in local observation.\\
It is clear that
the last arbitrary way of achieving the kind of invariance mentioned
is to require $\alpha(x)$ to be dimensionless.
Now, dimensional arguments may be used to
restrict the form of functional connection in (\ref{c2}) to
\begin{equation}
G_{ex.}^{-1}=\alpha(x) \frac{T_{mic.} M_{uni.}}{\hbar c^3},
\label{c3}\end{equation}
The order of magnitude of $\alpha(x)$
may be estimated by the requirement that
the intercession of $\alpha$ in the present epoch of the universe
would adjust $G_{ex.}^{-1}$ to a value close to the reciprocal
of the observed
value of gravitational constant.
This is a statement of numerical cooincidence of the
thermodynamically predicted value of $G$ with its observed value.
It gives the estimate
\begin{equation}
\alpha \sim 10^{-28}
\label{c4}\end{equation}
This relation has a significance in a rough-order-of magnitude manner
only, but
indicates the impossibility of achieving a cosmic
thermodynamic connection of $G$
with $\hbar$ and $c$ without producing an immensely small dimensionless
number.\\
Physically, there is one way of interpreting this
observation\footnote{The reader should notice the correspondence between
the next statement and the appreciation that the gravitational entropy
of the present universe is remarkably low [3].}: To an
observer restricted to observations in
the present epoch of the universe the allowed contribution
of the entire visible mass of the universe to the 'thermal' source of
gravitational entropy
appears to be remarkably
fractional, that is,
$\alpha$
may have the physical significance of an absolute scale factor which, if
applied to the entire visible mass of the universe, would provide a
measure of the 'entropic' mass scale of the universe,
$M_{ent.}$\footnote{If we take the gravitational entropy $s$ of
the expanding, time-asymmetric universe,
considered as
a thermodynamic system, as somehow related to $M_{uni.}$, we may associate
with the universe an entropic mass scale $M_{ent.}$ by
means of the relation (we are
using here the Planck units)
$$
\frac{ds}{dM_{uni.}}\sim M_{ent.}
$$
which, using the thermodynamic identity $ds/dM=1/T$ agrees
with (\ref{c3})
if and only if $M_{ent.}\sim \alpha M_{uni.}$. In this way $M_{ent.}$
may be pictured as the thermal source of gravitational entropy.}.
Of course, much work is needed to understand the dynamical origin of
the entropic mass scale of the universe\footnote{In a dynamical
approach it seems
reasonable to suspect a connection between $\alpha(x)$ and the
expectation value $<\phi^2>$ of a quantized scalar
field $\phi$. Such a theory, although conceptually different [4],
would have common
features with the Brance-Dicke theory [5].},
but in line with interpretation given
above, that mass scale
in a rough order of magnitude estimate should be
related to $M_{uni.}$ by a scale factor transformation applied to
$M_{uni.}$, namely
\begin{equation}
M_{ent.}~ \sim ~\alpha M_{uni.}.
\label{c5}\end{equation}
Since such a scale transformation
applied to the entire visible mass
of the universe may affect the dynamically ascertainable value of mass
for an arbitrarily localized and quasi isolated
gravitating system (considered as thermodynamic system)
in the cosmological context,
the appearance of
the entropic mass scale $M_{ent.}$ may prove to have
essential consequences.
First note that due to the very presence of $M_{ent.}$ it seems rather
natural to suspect
that the universe may be in a state in which
each localized and quasi isolated
gravitating system 'irreversibly' contributes to
$M_{ent.}$ by a fraction $\sim \alpha$ of
its mass. In this way $\alpha$ may
regulate
a sort of universal transfer of mass into the thermal source of
gravitational entropy.   \\
If a statement of this sort is admitted then,
due to the very existence of $M_{ent.}$, the
dynamically ascertainable value of mass for a
localized and quasi isolated gravitating system in
the cosmological context ought to exceed its 'bare' gravitational
mass $M$ in the 'idealized'
classical picture (the mass dynamically determined by the static
Schwarzschild-metric
at large distances from the system, see [6]) by a term of the order of
$\alpha M$.\\
For a black hole in
the cosmological context\footnote{There may be problems of principle
in the definition of a black hole in the cosmological context, for the
discussion of some aspects the reader is referred to [7].} formed by
an actual process of gravitational contraction
this has the remarkable
consequence that, asymptotically,
the temperature of the radiation would differ from Hawking's
temperature by a term of the order of $\alpha$, namely
\begin{equation}
T= T_H+~o~(\alpha),~~~
\alpha~ \sim~ 10^{-28}.
\label{c5}\end{equation}

\section{Scale-hierarchy}

There is now an important question regarding the effect of
cosmic thermalization of gravitational constant. Could it not be that
this thermalization could define an additional scale other then the linear
dimension of a black hole, i.e. its Schwarzschild-radius, for black hole
radiation?\\
Unavoidably, a new scale may be defined if we look for a characteristic
length scale, say $l$, for which the ratio of $l$ and the
linear dimension of a black hole, i.e. its Schwarzschild-radius $r_g$,
becomes of
the same order of magnitude as the small number
$\alpha$, namely
\begin{equation}
\frac{l}{r_g}~\sim \alpha.
\label{h1}\end{equation}
Now the relation (\ref{c5}) may be written as
\begin{equation}
T= T_H+~o~(\frac{l}{r_g}),
\label{h2}\end{equation}
which indicates that the real entropic state of a black hole
must unavoidably exhibit the scale hierarchy
\begin{equation}
l \ll r_g.
\label{h3}\end{equation}
Note that for a black hole of a typical astrophysical dimension
$\sim 10^5 cm$ the length $l$ will define a short distance scale
which in term of the Planck length $l_P$ would have the order of magnitude
$\sim 10^{10} l_P$.

\section{Short distance cut-off}

Recently there have been efforts in understanding the
role played by arbitrarily high frequencies apparently entering the usual
derivation of Hawking radiation [8][9][10]. The difficulty of
the usual derivation is that, given a static detector
placed far a way from a spherically symmetric black hole, the
long time response of the detector to outgoing modes of a quantized field
becomes (as a consequence of the 'infinite' gravitational redshift effect)
causally correlated to the behaviour of incoming modes with arbitrarily high
frequencies in the past, the relevant frequencies
increasing as (units are used in which $c=\hbar=G=1$)
\begin{equation}
\omega~\sim~\frac{1}{|v|}
\label{s1}\end{equation}
as the advanced time $v$ approaches the long time limit $v\rightarrow0$
($v=0$ corresponds to
the formation of the horizon), details may be found in [9].
It is now evident that in the long time
one comes very soon into the real difficulties
concerning the interpretation of 'infinitely' magnified
frequencies.
I would like to argue that the difficulty might have its origin
in an improper separation of the entropic state of a black hole
from the effect of scale-hierarchy (\ref{h3}). I am very well a ware
of the difficulty of the problem, still the following extremely heuristic
remark appears to be worth mentioning.\\
The point is that the horizon, which in the usual
analysis forms at $v=0$, in the presence of $l$ is expected to form
at a slightly smaller value of the advanced time
(since the scale
$l$ has its origin in an increment of the black hole mass). This
'entropic' shifting of horizon
may have the following consequence. First, note that
in the frame of a free falling observer
beginning his journey from the rest at infinity the advanced time
near the horizon would change oppositly at a rate which is of the order
of the rate at which the Schwarzschild-radial coordinate $r$
would change, see appendix. Thus, in the presence of $l$
the entropic shifting of horizon may
give rise to the possibility of the existence of an effective cut-off
frequency $\sim~l^{-1}$ for incoming modes in that frame.
The heuristic nature of this argument, however, should
once again be stressed.

\section{Summary and Outlook}

The essence of the investigations made above is that the
real entropic state of a black hole can
not be separated from the state of the universe as a whole. This aspect is
clearly reflected in the presence of the
short distance scale $l$ in the scale hierarchy (\ref{h3}).
This scale
has its origin in a remarkably fractional contribution of the mass of
a black hole (in the moment of its formation)
to the entropic mass scale of the universe and may have the effect
of prohibiting the unphysical ultrahigh frequency modes apparently
entering the usual derivation of Hawking's radiation.\\
There are many open problems to be resolved. For example, due to
the 'infinite' gravitational redshift effect, the causal structure
of black hole space-time in the semiclassical picture appears
to be very sensitive to an (even)
small entropic shifting in the location of horizon,
leading, therefore, to
the problem of how the appearance of $l$ may limit the validity  of the
conclusions drawn in the semiclassical picture. In addition
there is the problem of
properly understanding the physical
status of the length scale $l$. Regarding this last problem
it would seem, first, that
the actual process of gravitational contraction of a sufficiently
large mass in the present epoch of the universe may define a
minimal bound for physical length scales as $l$ in (\ref{h3}) appears
to be 'entropically' bounded from below  by the limit mass of the
hydrostatic stability for a super dense star. It is quite conceivable
that the appearance of this 'entropic' minimal length may act as a possible
reason for the postulate of a fundamental irreversibility
inherent in ultrashort distance physics and for the break
down of the universal requirement of Lorentz symmetry.

\vskip 2cm

{\bf Appendix} \\
Let $p$ be an affin parameter along a free falling geodesics of the
Schwarzschild-metric. Since this metric does not depend on the
Schwarzschild-time $t$ the quantity $p_t=g_{t\mu}dx^{\mu}/dp=
(1-\frac{2M}{r})dt/dp$ is conserved. Thus, the affin parameter may
normalized so that
\begin{equation}
\frac{dt}{dp}=(1-\frac{2M}{r}).
\label{AP1}\end{equation}
In this normalization the equation governing the radial motion of
the geodesics takes the form
\begin{equation}
(\frac{dr}{dp})^2+(1-\frac{2M}{r})(\frac{J^2}{r^2}+E)-1=0,
\label{AP2}\end{equation}
where $E$and $J$ are two integration constants, the first one being related
to the proper time by, see [11]
\begin{equation}
d\tau^2=E dp^2.
\label{AP3}\end{equation}
If the geodesics starts from the rest at infinity, it follows from
(\ref{AP2}) $E=1$,
and consequently (\ref{AP2}) with respect to $\tau$ can be written as
\begin{equation}
(\frac{dr}{d\tau})^2+(1-\frac{2M}{r})(\frac{J^2}{r^2}+1)-1=0.
\label{AP4}\end{equation}
Near the horizon $r\approx2M$, we may write it in the form
\begin{equation}
(\frac{dr}{d\tau})^2-\frac{2M}{r} \approx 0,
\label{AP5}\end{equation}
Thus, the rate of the change of $r$ with respect to the proper
time near the horizon must be $dr/d\tau\approx -(2M/r)^{1/2}$. Now, it is
easy to determine the rate at which advanced time $v$ would change near
the horizon along the geodesics. Using
\begin{equation}
v=t+\stackrel{*}{r},~~~\stackrel{*}{r}=r+2M \ln(\frac{r}{2M}-1),
\label{AP6}\end{equation}
one finds $dv/d\tau \approx 1/2$.
\\\\

{\bf References}\\\\
\begin{tabular}{r p{12cm}}
1. & Bekenstein J D 1973 Phys. Rev. D7,2333  \\
2. & Hawking S W 1975 Commun. Math. Phys. 43 \\
3. & Penrose R 1979 General Relativity, an Einstein Centenary Survay ed.
by Hawking S W and Israel W, p.581 (Cambridge university Press)\\
4. & Salehi H Proceeding of international
symposium of generalized symmetries, Germany, World Scientific (1994)\\
5. & Brance C and Dicke R 1961 Phys. Rev. D 124, 3, 925\\
6. & Tolman Relativity Thermodynamics and Cosmology Oxford (1966)\\
7. & Qadir A Proc. of the fifth. Marcell Grossman Meeting on general
relativity, Part A. World Sceitific (1989)\\
8. & Jacobson Th 1991 Phys. Rev. D 44,1731 \\
9. & Salehi H 1992 Class. Quantum Grav.9 2557-2571\\
10 & Jacobson Th 1993 Phys. Rev. D 48,728 \\
11. & Weinberg S Gravitation and Cosmology, John Wiley and Sons (1972)\\
\end{tabular}

\end{document}